\documentclass[%
 jap,
 jmp,%
 amsmath,amssymb,
 reprint,%
]{revtex4-1}

\usepackage{graphicx}
\usepackage{dcolumn}
\usepackage{bm}
\usepackage{amsmath}

\begin{document}





\title{Tuneable plasmonics enabled by capillary oscillations of liquid-metal nanodroplets}
\author{Ivan S. Maksymov and Andrew D. Greentree}
\affiliation{ARC Centre of Excellence for Nanoscale BioPhotonics, School of Science, RMIT University, Melbourne, VIC 3001, Australia}

\date{\today}

\begin{abstract}
Plasmonics allows manipulating light at the nanoscale, but has limitations due to the static nature of nanostructures and lack of tuneability. We propose and theoretically analyse a room-temperature liquid-metal nanodroplet that changes its shape, and therefore tunes the plasmon resonance frequency, due to capillary oscillations. We show the possibility to tune the capillary oscillation frequency of the nanodroplet and to drive the oscillations electrically or mechanically. Employed as a tuneable nanoantenna, the nanodroplet may find applications in sensors, imaging, microscopy, and medicine. 
\end{abstract}


\maketitle 


Nanoplasmonics paves the way for controlling light at the subwavelength scale \cite{Klimov}. However, once a metal nanostructure has been fabricated, its optical characteristics cannot be changed in a reversible manner, which restricts its potential applications and does not allow meeting an increasing demand for tunable optical properties. Therefore, a large and growing body of research investigates tuneable plasmonic structures such as, e.g., optical nanoantennae \cite{Nov11, Agi13}. Spectral tuneability of nanoantennae has been demonstrated by combining a metal nanostructure with functional materials such as liquid crystals, metamaterials, elastomers, semiconductors, phase-changing media, nonlinear and magneto-optical materials (for a review see, e.g., \cite{Mak12, Agi13, Mak16}). Because of many options in size, material and features, nanoantennae have been employed in tuneable filters, sensors, switches, on-chip optical links, sources of quantum light, etc., and found applications in imaging, spectroscopy, microscopy, photovoltaics, and medicine \cite{Mak12, Agi13, Mak16}.

In this paper, we propose and theoretically analyse a room-temperature liquid-metal nanodroplet that changes its shape, and therefore tunes the plasmon resonance frequency, due to capillary oscillations. Capillary oscillations of the nanodroplet arise because of a competition between inertia and surface tension \cite{Rayleigh}, and may be driven electrically or mechanically (see, e.g., \cite{Oh08}). In our study, for simplicity and to highlight important physics, the nanodroplet oscillates in a vacuum or air. As its constituent material we consider a room-temperature liquid-gallium metal alloy that, in contrast to liquid mercury, is non-toxic and has a relatively low density \cite{Mor08}. We demonstrate that, similar to solid-gallium nanoparticles of $\sim 100-300$~nm diameter \cite{Kni15}, liquid-gallium nanodroplets of comparable size have plasmon resonance frequencies in the ultraviolet spectral range. We also show that the good mechanical properties of liquid-gallium alloys, such as a large and reversibly changeable surface tension and low viscosity \cite{Mor08, Kha14}, enable capillary oscillations with a low damping rate and oscillation frequency tuneable in the MHz-to-GHz range. By considering the scenario of capillary oscillations driven electrically or mechanically, we calculate that spectral tuning of the plasmon resonance in the $150-300$~nm spectral range is possible with experimentally achievable peak amplitudes of the applied ac voltage or mechanical pressure.

Thus, liquid-metal nanodroplets may operate as tuneable optical nanoantennae and therefore have the same wide range of applications in photonic devices conceivable with nanoantennae \cite{Nov11, Mak12, Agi13, Mak16}. We also envision applications in emergent areas such as sensing of sound at the nanoscale \cite{Ma16, Mak16_SciRep}, unconventional photonics \cite{Mak17_1, Mak17_2}, and detection of mechanical signatures of cells and bacteria \cite{Zin09}. Finally, although our analysis is mostly focused on nanodroplets oscillating in a vacuum or air, our findings are extendable to the case of liquid-metal nanodroplets immersed in a liquid \cite{Lu15} where capillary oscillations may be driven by ultrasound.

\begin{figure*}[t]
\centering\includegraphics[width=16cm]{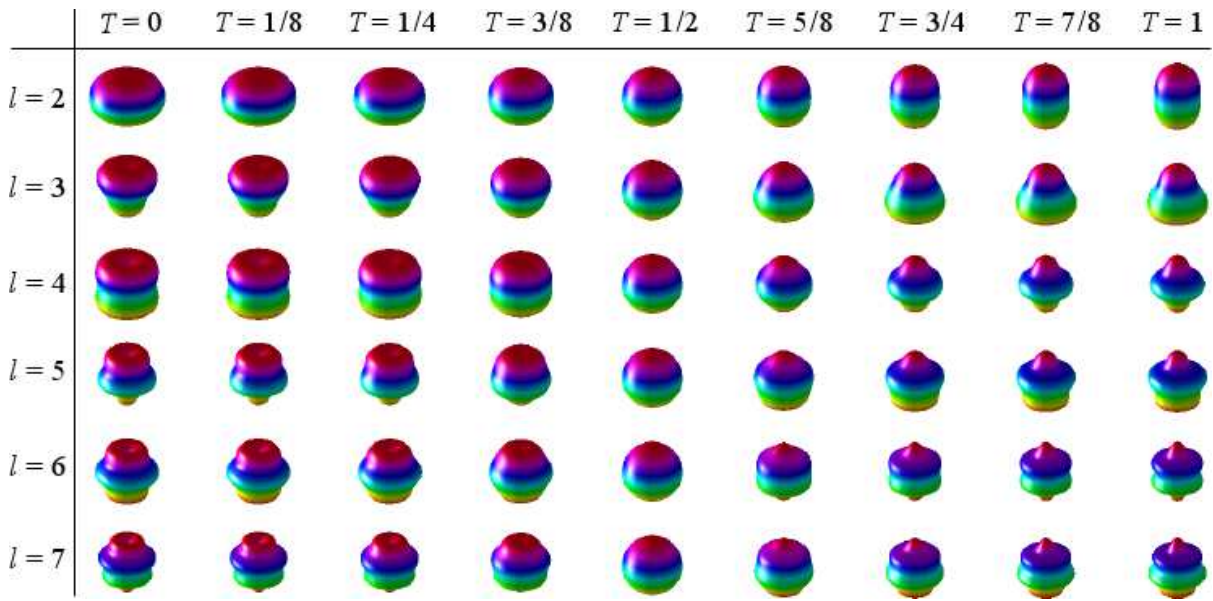}
\caption{Representative oscillation mode shapes of plasmonic liquid-gallium alloy nanodroplets through a half-period of oscillation for the $l=2..7$ modes. The peak oscillation amplitude for all modes is $A_{l}/R=0.3$ as justified in the main text. $T_{l}=\omega_{l}t$ is given in the units of $\pi$ radians ($T_{l}=1$ is a half of period).}
\label{fig1}
\end{figure*}

Applications of microscopic liquid droplets in photonics were previously discussed in \cite{Cro13, Maa16, Dah16}. For example, in \cite{Maa16} a water-droplet microresonator fabricated on the edge of an optical fibre was demonstrated. However, a large size and low optical refractive index ($n=1.33$ for water) of liquid droplets prevent them from applications at the nanoscale. Moreover, preference has mostly been given to acoustic-like oscillations (MHz-range frequencies) of microscopic liquid droplets rather than to capillary ones (kHz range) because of the focus on applications in high-frequency opto-mechanics \cite{Dah16}. 

However, the nanoscale size of liquid-metal droplets in combination with a high surface tension $\sigma$ and relatively low density $\rho$ of liquid-gallium alloys ($\sim 10$ and $\sim 6$ times $\sigma$ and $\rho$ of water, respectively \cite{Mor08}) allows achieving capillary oscillations in the GHz-frequency range. Thus, we consider capillary oscillations of a single, initially spherical liquid-gallium alloy nanodroplet and analyse the resonance angular frequency $\omega_{l}$ and peak amplitude $A_{l}$ of the capillary oscillation modes with the mode numbers $l$.

In the linear approximation, $\omega_{l}$ of a nanodroplet oscillating in a vacuum or air is given by \cite{Rayleigh}

\begin{align}
\omega^{2}_{l} = l(l-1)(l+2)\sigma/(\rho R^{3})
\label{eq:one},
\end{align}

\noindent where $\sigma$ and $\rho$ are the surface tension of the liquid metal and density, respectively. The modes $l=0$ and $l=1$ are zero-frequency modes corresponding to conservation of volume and translational invariance, respectively. The lowest non-zero frequency (fundamental) mode excited in experimental conditions is $l=2$. 

The $3$D shapes of the oscillation modes are given by 

\begin{align}
r = R \left[1 + (A_{l}/R) \cos(\omega_{l} t) P_{l} (\cos \theta) \right]
\label{eq:two},
\end{align}   

\noindent where the coordinate origin is at the centre of the nanodroplet, $A_{l}/R$ is the peak amplitude of the $l$th mode, $t$ is time, $\theta = 0...2 \pi$, and $P_{l}(x) = \sum_{m=0}^{M} (-1)^{m}\frac{(2l-2m)!}{2^{l}m!(l-m)!(l-2m)!} x^{l-2m}$ where $M=l/2$ or $M=(l-1)/2$ whichever is an integer. 

Figure~\ref{fig1} shows representative oscillation mode shapes through a half-period of oscillation for the modes $l=2...7$ with the peak oscillation amplitude $A_{l}/R=0.3$. Within the half-period of oscillation the nanodroplet assumes the shapes at $T=0$ and $T=1$ corresponding to the largest deviation from the spherical shape $T=1/2$ (here $T_{l}=\omega_{l}t$, the index $l$ is omitted for simplicity). For the fundamental $l=2$ mode the shape of the nanodroplet changes from an oblate to prolate spheroid. For $l=3$ it changes from an inverted pyramid to a pyramid, and so on. For the second half of the period the nanodroplet retraces the shapes assumed in the fist half.

In general, Eq.~\ref{eq:one} and Eq.~\ref{eq:two} are valid for linear capillary oscillations of an inviscid and incompressible liquid droplet with the infinitesimal amplitudes $A_{l}/R \rightarrow 0$ \cite{Rayleigh}. This linear theory was developed by Rayleigh and later extended to take into account viscosity of the droplet and consider the scenario of a viscous liquid droplet immersed in another viscous liquid \cite{Chandrasekhar, Mil68}.

Small amplitude ($A_{l}/R \approx 0.1...0.4$) oscillations were analysed in \cite{Tsa83} and a decrease in $\omega_{l}$ was predicted for the modes $l=2...4$ as $\tilde{\omega}_{l}=\omega_{l}[1-\gamma_{l} (A_{l}/R)^{2}]$. The values of $\gamma_{l}$ can be found in \cite{Tsa83}.

Large amplitude ($A_{l}/R \gtrsim 0.4$) nonlinear oscillations of inviscid droplets were investigated in \cite{Pat91}, and viscosity was taken into account in \cite{Bas92}. The numerical method proposed in \cite{Bas92} also works in the case of small amplitude oscillations and, together with earlier numerical results \cite{Foote, Alonso}, it confirms the accuracy of the theory from \cite{Tsa83}.

Using the corrections for the oscillation mode shapes from \cite{Tsa83}, we analyse the difference between the predictions of the linear and nonlinear theories. At the peak oscillation amplitudes $A_{l}/R = 0.4$ considered in \cite{Tsa83, Foote, Alonso} for the modes $l=2...4$, the variations in the shape of the oscillating droplet follow the predictions of the linear theory within $\sim 10\%$ accuracy, with the maximum deviation occurring near the equator and the poles of the droplet. The most notable difference is observed when the linear theory predicts the return of the droplet to a perfectly spherical shape, but the nonlinear theory produces a prolate spheroid shape for the fundamental mode ($l=2$) and multilobed shapes for the modes $l=3$ and $l=4$.

Consequently, we expect a $<10\%$ difference between the oscillation mode shapes in Fig.~\ref{eq:one} obtained in the linear approximation and the predictions of the nonlinear theory and simulations in \cite{Tsa83, Foote, Alonso}, because the peak amplitude $A_{l}/R = 0.3$ used in our analysis is smaller than that in those papers. As in this work we aim to demonstrate spectral tuning in the optical domain, this difference will not qualitatively affect our discussion. Furthermore, the use of a simple linear theory will be of great help for establishing a relationship between the optical tuning and the force that drives the capillary oscillations. Thus, in the following we rely on the linear theory only.

\begin{figure}[t]
\centering\includegraphics[width=6.5cm]{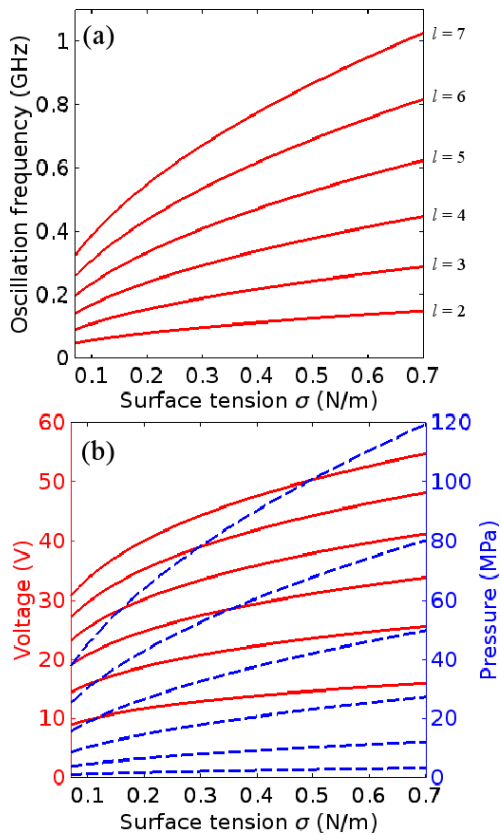}
\caption{(a) Resonance oscillation frequency of the modes $l=2...7$ as a function of the surface tension $\sigma$. (b) Voltage (left y-axis, solid curves) and pressure (right y-axis, dashed curves) required to achieve the peak amplitudes $A_{l}/R=0.3$ for the modes $l=2...7$ as a function of $\sigma$. Similar to panel (a), the lowest (highest) curves correspond to the mode $l=2$ ($l=7$).}
\label{fig2}
\end{figure}

We analyse the shape mode equation \cite{Rayleigh, Chandrasekhar, Oh08} to define the strength of the oscillation driving force $F_{l}$ required to produce the modal shapes in Fig.~\ref{fig1} 

\begingroup\makeatletter\def\f@size{9}\check@mathfonts
\def\maketag@@@#1{\hbox{\m@th\large\normalfont#1}}%
\begin{align}
\ddot a_{l} + \frac{2(l-1)(2l+1) \mu}{\rho R^{2}} \dot a_{l} + \frac{l(l-1)(l+2)\sigma}{\rho R^{3}} a_{l} = \frac{lF_{l}}{\rho R}
\label{eq:three},
\end{align} \endgroup

\noindent where $\omega_{l}$ is given by Eq.~\ref{eq:one}, $\mu$ is the viscosity of the liquid metal, and $a_{l}$ is a function of time $t$ representing the instantaneous amplitude of the $l$th oscillation mode. For periodic $F_{l}$, Eq.~\ref{eq:three} describes a periodically forced damped harmonic oscillator, for which the general solution is the sum of a transient solution and a steady-state response.

We are interested in the amplitude of the steady-state response because it relates the value of $A_{l}/R$ to the peak amplitude of $F_{l}$. As the oscillation force we consider mechanical pressure or applied ac voltage. In the case of a nanodroplet immersed in another liquid the former also applies to ultrasound, but the latter models the scenario of electrowetting -- the electrical control of wettability used to handle nanodroplets \cite{Oh08}. 

A typical electrowetting setup employs a thin insulating layer (thickness $d$ and radio-frequency range dielectric permittivity $\epsilon_{\rm{d}}$) that separates the droplet from the counter electrode. According to \cite{Oh08}, for electrowetting $F_{l}=(2l+1)\epsilon_{\rm{0}} \epsilon_{\rm{d}} V_{l}^{2}/(2dR)$, where $\epsilon_{\rm{0}}$ is the vacuum permittivity and $V_{l}$ is the peak amplitude of the applied ac voltage for the $l$th mode. However, because in our model the nanodroplet is not attached to a surface, in calculations we use the generic values $d=100$~nm and $\epsilon_{\rm{d}}=6$, which may be suitable for design of real-life devices.  

At the resonance, for the peak amplitude $A_{l}/R=0.3$ from the steady-state solution to Eq.~\ref{eq:three} we obtain $F_{l} = 0.3R^{2} b \omega_{l} \rho/l$ for the mechanical pressure excitation and $V_{l} = \sqrt{0.6R^{3} b \omega_{l} \rho d/[l(2l+1) \epsilon_{\rm{0}} \epsilon_{\rm{d}}]}$ for the ac voltage excitation. Here $b=2(l-1)(2l+1)\mu/(\rho R^2)$,

We use typical liquid-gallium alloy parameters \cite{Mor08, Kha14}: density $\rho = 6360$~kg/m$^3$, viscosity $\mu=0.0024$~Pa~s, and surface tension reversibly changeable from $\sigma=0.7$~N/m down to $0.07$~N/m (approximately $\sigma$ of water) \cite{Kha14}. The radius of the spherical (undeformed) nanodroplet $R = 100$~nm is chosen because solid-gallium nanoparticles of comparable size support plasmon resonances in the ultraviolet spectral range, in which the dielectric functions of liquid and solid gallium are nearly identical \cite{Kni15}.

Figure~\ref{eq:two}(a) shows the resonant oscillation frequency $f_{l}=\frac{\omega_{l}}{2 \pi}$ as a function of the surface tension $\sigma$ for the modes $l=2...7$. Figure~\ref{eq:two}(b) shows the voltage (left y-axis, solid curves) and pressure (right y-axis, dashed curves) required to achieve the peak amplitudes $A_{l}/R=0.3$ for the modes $l=2..7$, as a function of $\sigma$. We observe that the lowest possible oscillation frequency is $\sim 40$~MHz for the mode $l=2$, but for the mode $l=7$ it can reach $1$~GHz. The applied voltage (pressure) required to achieve $A_{l}/R=0.3$ at these frequencies ranges from $\sim10$~V ($\sim 4$~MPa) to $\sim 55$~V ($120$~MPa), which are achievable in experiment values \cite{Oh08, Mak17_1}.

\begin{figure}[t]
\centering\includegraphics[width=8.5cm]{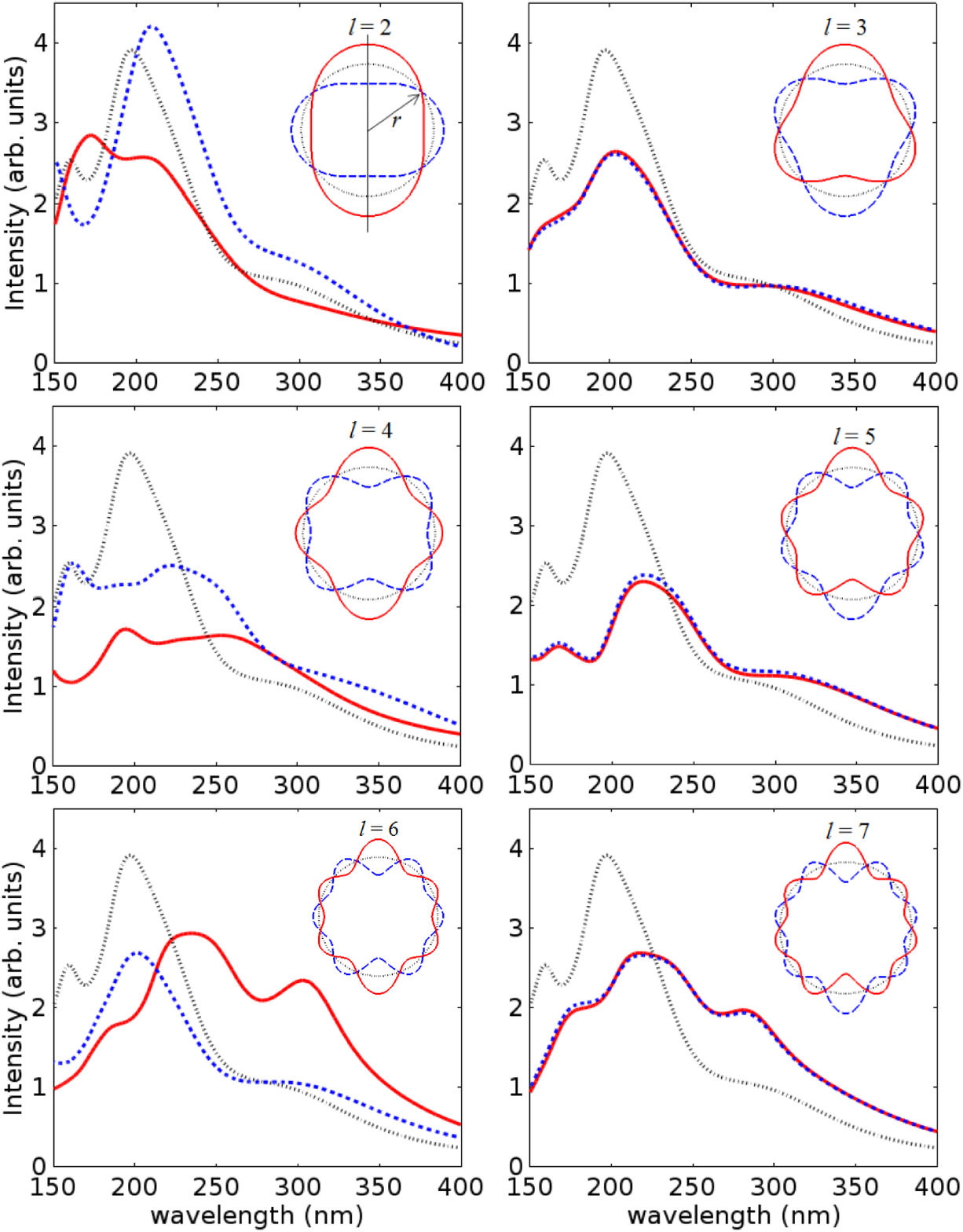}
\caption{Optical intensity spectra calculated for the static $3$D shapes assumed by the nanodroplet at $T=0$ (dashed curves), $T=1$ (solid curves), and $T=1/2$ (dotted curve) shown in Fig.~\ref{eq:one}. The insets show the cross-sections of these $3$D shapes. The same curve styles are used in the insets and the main panels.}
\label{fig3}
\end{figure}  

Figure~\ref{fig3} shows the calculated optical intensity spectra of the nanodroplet at $T=0$, $T=1$, and $T=1/2$ (spherical shape) for the modes $l=2...7$. Three-dimensional ($3$D), static shapes of the nanodroplet from Fig.~\ref{fig1} were used in the calculations. The the insets in Fig.~\ref{fig3} show the cross-sections of the $3$D shapes at $T=0$, $T=1$, and $T=1/2$ calculated using Eq.~\ref{eq:two}. The optical  spectra were calculated using a $3$D optical Finite-Difference Time-Domain (FDTD) method with the spatial discretisation $2$~nm (see \cite{Mak16_SciRep} for details). We use the dielectric permittivity function of liquid gallium \cite{Kni15}, which should be a good model for gallium-based alloys. In calculations, we ignore the presence of a nanometres-thin native gallium oxide layer formed on top of the liquid-gallium alloy because this layer is easily removable \cite{Kha14}. The calculated intensity is normalised to the intensity of the incident light such that the values $>1$ indicate a field enhancement due to the plasmon resonance.

We observe the spectral tuning in the $150-300$~nm range due to the shape deformation of the nanodroplet. In particular, as compared with the spherical shape ($T=1/2$), the plasmon resonance peaks shift in frequency and their intensity is changed. Moreover, additional resonance peaks appear in the spectra for the higher-order oscillation modes. 

For the modes $l=2,4,6$ the spectra produced by the modal shapes at $T=0$ and $T=1$ are different, but for the modes $l=3,5,7$ they coincide to almost graphical accuracy. Hence, the incident light senses more variations in the nanodroplet's shape tuned on one of its even modes than when an odd mode is excited. This result is attributed to the symmetry (asymmetry) of the even (odd) modes with respect to the horizontal axis in the insets in Fig.~\ref{fig3}. Whereas the deformed nanodroplet shapes for, e.g., the $l=3$ mode are the same but rotated by $\pi$ radians, for the mode $l=4$ they are different. This results in different optical properties.

In conclusion, we have proposed and theoretically analysed a room-temperature liquid-metal nanodroplet as a spectrally tuneable plasmonic structure, which may be used as a tuneable optical nanoantenna having applications in nanoscale optical filters, sensors, opto-mechanic systems, and transducers of electrical and mechanical forces into optical signals (e.g., nanoscale microphones). The ability to tune the plasmon resonances originates from capillary oscillations of the nanodroplets, which may have the peak amplitudes of $\sim 30\%$ of the undeformed radius at attainable in experiments levels of the oscillation driving force such as applied ac voltage of $\sim 10-50$~V or pressure of $\sim 5-100$~MPa. It is noteworthy that the predicted large changes in the shape of the liquid-metal nanodroplets cannot be achieved in solid-metal nanoparticles \cite{Rui12} due to mechanical hardness of the latter. Although oscillations with comparable amplitudes are possible with gas nanobubbles in liquids \cite{Mak17_2}, nanobubbles have a low refractive index contrast with the host liquid ($\Delta n \approx 0.33$ in water) and thus require auxiliary plasmonic nanostructures to enable their interaction with the incident light \cite{Mak17_2}. In contrast, liquid-metal nanodroplets combine both acoustic and optical properties.

Furthermore, the ability to manipulate the surface tension of liquid-gallium metal alloys \cite{Kha14} opens up opportunities to tune the oscillation frequency from several MHz to several GHz. This is difficult to achieve in solid-metal nanoparticles that vibrate mostly at frequencies $>1$~GHz and these frequencies cannot be tuned. Finally, liquid-gallium alloy nanodroplets may be frozen at temperatures $<10-15^{\rm{o}}$C or so \cite{Mor08}, which converts them into solid-state nanoparticles with virtually the same plasmonic properties \cite{Kni15} but having one of the complex $3$D shapes shown in Fig.~\ref{fig1}. Thus, variations of the temperature around the freezing point of the liquid-gallium alloy may be used to switch the nanodroplet from the liquid to solid phase and vice-versa. However, because liquid-gallium alloys tend to supercool below the freezing point, the liquid nanodroplet has to be perturbed to create crystal nucleation sites required to initiate freezing.

This work was supported by the Australian Research Council (ARC) through its Centre of Excellence for Nanoscale BioPhotonics (CE140100003), LIEF program (LE160100051) and Future Fellowship (FT1600357). This research was undertaken on the NCI National Facility in Canberra, Australia, which is supported by the Australian Commonwealth Government.



\begin{thebibliography}{99}

\bibitem{Klimov} V. Klimov, \textit{Nanoplasmonics} (Pan Stanford, Singapore, 2014).

\bibitem{Nov11} L. Novotny and N. van Hulst, Nature Photon. {\bf 5,} 83 (2011).

\bibitem{Agi13} M. Agio and A. Al\`{u}, \textit{Optical Antennas} (Cambridge University Press, Cambridge, 2013).

\bibitem{Mak12} I. S. Maksymov, I. Staude, A. E. Miroshnichenko, and Yu. S. Kivshar, Nanophotonics \textbf{1,} 65 (2012).

\bibitem{Mak16} I. S. Maksymov, Rev. Phys. \textbf{1,} 36 (2016).

\bibitem{Rayleigh} Lord Rayleigh, Proc. R. Soc. Lond. \textbf{29,} 71 (1879).

\bibitem{Oh08} J. M. Oh, S. H. Ko, and K. H. Kang, Langmuir \textbf{24,} 8379 (2008).

\bibitem{Mor08} N. B. Morley, J. Burris, L. C. Cadwallader, and M. D. Nornberg, Rev. Sci. Instrum. \textbf{79,} 056107 (2008).

\bibitem{Kni15} M. W. Knight, T. Coenen, Y. Yang, B. J. M. Brenny, M. Losurdo, A. S. Brown, H. O. Everitt, and A. Polman, ACS Nano \textbf{9,} 2049 (2015).

\bibitem{Kha14} M. R. Khan, C. B. Eaker, E. F. Bowden, and M. D. Dickey, PNAS \textbf{111,} 14047 (2014).

\bibitem{Mak16_SciRep} I. S. Maksymov and A. D. Greentree, Sci. Rep. \textbf{6}, 32892 (2016).

\bibitem{Ma16} Y. Ma, Q. Huang, T. Li, J. Villanueva, N. H. Nguyen, J. Friend, and D. J. Sirbuly, ACS Photonics \textbf{3}, 1762 (2016).

\bibitem{Mak17_1} I. S. Maksymov and A. D. Greentree, Opt. Express \textbf{25,} 7496 (2017).

\bibitem{Mak17_2} I. S. Maksymov and A. D. Greentree, Phys. Rev. A \textbf{95,} 033811 (2017).

\bibitem{Zin09} P. V. Zinin and J. S. Allen, III, Phys. Rev. E \textbf{79,} 021910 (2009).

\bibitem{Lu15} Y. Lu, Q. Hu, Y. Lin, D. B. Pacardo, C. Wang, W. Sun, F. S. Ligler, M. D. Dickey, and Z. Gu, Nat. Commun. \textbf{6,} 10066 (2015).

\bibitem{Cro13} Th. Crouzil and M. Perrin, J. Europ. Opt. Soc. Rap. Public. \textbf{8,} 13079 (2013).

\bibitem{Dah16} R. Dahan, L. L. Martin, and T. Carmon, Optica {\bf 3,} 175 (2016).

\bibitem{Maa16} S. Maayani, L. L. Martin, S. Kaminski, and T. Carmon, Optica \textbf{3,} 552 (2016).

\bibitem{Chandrasekhar} S. Chandrasekhar, \textit{Hydrodynamic and Hydromagnetic Stability} (Oxford University Press, Oxford, 1961).

\bibitem{Mil68} C. Miller and L. Scriven, J. Fluid Mech. \textbf{32,} 417 (1968).

\bibitem{Tsa83} J. A. Tsamopoulos and R. A. Brown, J. Fluid Mech. \textbf{127,} 519 (1983).

\bibitem{Pat91} T. W. Patzek, R. E. Benner, Jr, O. Basaran, and L. E. Scriven, J. Comp. Phys. \textbf{97,} 489 (1991).

\bibitem{Bas92} O. Basaran, J. Fluid Mech. \textbf{241,} 169 (1992).

\bibitem{Foote} G. B. Foote, J. Comp. Phys. \textbf{11,} 507 (1973).

\bibitem{Alonso} C. T. Alonso, JPL Proc. of the Intern. Colloq. on Drops and Bubbles \textbf{1,} 139 (1974).

\bibitem{Rui12} P. V. Ruijgrok, P. Zijlstra, A. L. Tchebotareva, and M. Orrit, Nano Lett. \textbf{12,} 1063 (2012).


\end{thebibliography}
\end{document}